\documentclass[prl,showpacs,twocolumn]{revtex4}

\usepackage[]{amsmath,amssymb,amsfonts}
\usepackage[]{amsthm}
\usepackage[]{graphicx}

\renewcommand{\Re}{\mathrm{Re}}
\newcommand{\Ma}{\mathrm{Ma}}

\newcommand{\g}{\gamma}
\renewcommand{\r}{\rho}
\newcommand{\s}{\sigma}
\newcommand{\D}{\Delta}

\usepackage{color}
\usepackage[normalem]{ulem}

\begin{document} 


\title{Coherent structures in Dissipative Particle Dynamics simulations of the transition to turbulence in compressible shear flows}

\author{Jan-Willem \surname{van de Meent}}
\author{Alexander Morozov}\thanks{Present address: School of Physics, University of Edinburgh, Edinburgh EH9 3JZ, Scotland}
\author{Ell\'ak Somfai}\thanks{Present address: Department of Physics and Centre for Complexity Science, University of Warwick, Coventry, CV4 7AL, UK}
\author{Eric Sultan}
\author{Wim \surname{van Saarloos}}
\affiliation{Instituut-Lorentz, University of Leiden, Postbus 9506, 2300 RA Leiden, The Netherlands}


\begin{abstract}
  We present simulations of coherent structures in compressible flows near the transition to turbulence using the Dissipative Particle Dynamics (DPD) method. The structures we find are remarkably consistent with experimental observations and DNS simulations of incompressible flows, despite a difference in Mach number of several orders of magnitude. The bifurcation from the laminar flow is bistable and shifts to higher Reynolds numbers when the fluid becomes more compressible. This work underlines the robustness of coherent structures in the transition to turbulence and illustrates the ability of particle-based methods to reproduce complex non-linear instabilities.
\end{abstract}

\maketitle

\begin{figure}[t]
 \includegraphics[width=3.4in]{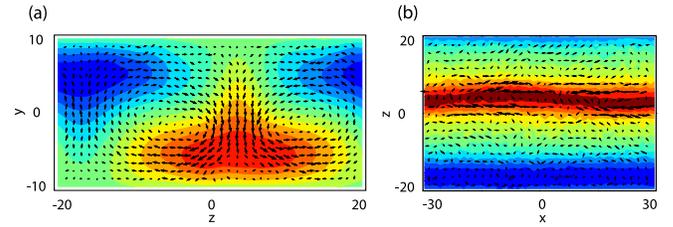}
 \caption{\label{fig:couette_2d} A pair of streaks in compressible plane Couette flow at $\Re = 1400$ and $\Ma=8$. Color contours denote the streamwise deviation from the laminar flow, vector fields denote the average in-plane motion. (a) Streamwise-direction view, showing $x$-averaged velocities. (b) Flow gradient-direction view at $y=0$. The fast streak shows a clear sinusoidal inflection.
 }
\end{figure}

The transition to turbulence in parallel shear flows like pressure-driven channel flow or flow in a pipe is one of the classic problems of fluid mechanics. Until recently even predicting the correct order of magnitude for the transitional Reynolds number $\Re$ was problematic. This situation has changed with the discovery of exact nonlinear solutions of the Navier-Stokes equations \cite{nagata_jfm_1990,waleffe_prl_1998,waleffe_pf_2003,faisst_prl_2003,wedin_jfm_2004,wang_prl_2007}. These solutions are dominated by streaks and streamwise vortices -- low-dimensional coherent structures observed experimentally in wall-shear flows \cite{robinson_arfm_1991} which are generated via the \emph{self-sustaining process} (SSP) proposed by Waleffe \cite{waleffe_pf_1997}. In the SSP, the counter-rotating quasi-streamwise vortices redistribute momentum along the wall-normal axis, creating spanwise modulations of the streamwise velocity, known as \emph{streaks}. The streaks in turn are subject to a Kelvin-Helmholtz-like instability due to the large velocity gradient across their surface. Nonlinear interactions between the instability modes couple back to the original streamwise vortices thus closing the cycle. Even though the exact solutions are linearly unstable, they have been shown to control the transition to turbulence and turbulent dynamics at moderate $\Re$ \cite{hof_science_2004,hof_prl_2005,skufca_prl_2006,kerswell_jfm_2007,wang_prl_2007,viswanath_archive_2007,gibson_archive_2007}. This scenario \cite{kerswell_nonlinearity_2005,eckhardt_arfm_2007} has emerged from a combination of intricate experiments \cite{hof_science_2004,hof_prl_2005}, large-scale numerical studies \cite{nagata_jfm_1990,waleffe_prl_1998,waleffe_pf_2003,faisst_prl_2003,wedin_jfm_2004,wang_prl_2007,skufca_prl_2006,kerswell_jfm_2007,viswanath_archive_2007,gibson_archive_2007} and  model equations studies \cite{waleffe_pf_1997,moehlis_njp_2004}. 

In this Letter we study the robustness of the coherent structures and the self-sustaining process at the onset of turbulence in compressible flows using a particle based method, the so-called Dissipative Particle Dynamics or DPD simulation method \cite{hoogerbrugge_epl_1992,groot_jcp_1997}. Such a simulation method represents a fluid by discrete interacting particles whose motion converges to hydrodynamic behavior on length-scales larger than the typical interparticle distance. Our results not only illustrate  the surprising robustness of the coherent structures \cite{robinson_arfm_1991,kerswell_nonlinearity_2005,eckhardt_arfm_2007} to thermal fluctuations and compressibility effects, but also show that studies of the transition to turbulence provide a very attractive and informative testbed for assessing the strengths and weaknesses of particle based simulation methods. Lattice-Boltzmann methods \cite{succi_2001} have already been successfully applied in supercomputer turbulence studies, like flow past a car \cite{chen_sci_2003}; we demonstrate here that the transition to turbulence can be studied effectively on a regular single node computer and that the DPD results throw new light on the robustness of the transition mechanism as well as on the simulation method itself.

\begin{figure*}[!t]
  \includegraphics[width=6.8in]{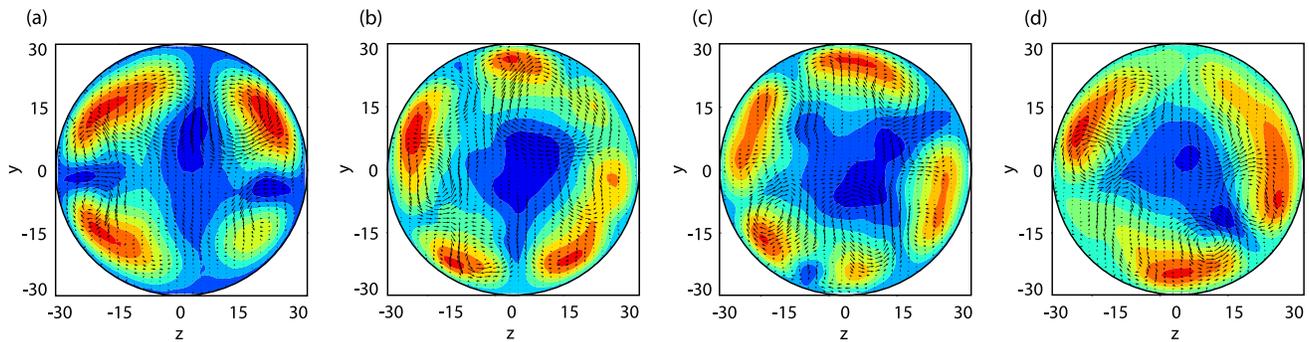}
  \vspace*{-3mm}
  
 \caption{\label{fig:pipe_2d} Snapshots of streaky pipe flow averaged over the pipe length. The simulation box has pipe diameter $d=60$ and periodic length $l=126$. DPD parameters are $(a,\g,kT)=(5,1,0.2)$, density $\rho=4$, $N=1425026$, yielding $\nu=0.2$, $c \simeq 4.2$ and $\Re/\Ma=1260$. (a) a 4-streak configuration at $\Re=1700$. (b)-(d) three different snapshots at $\Re=2700$. These states and the stochastic hopping between them are similar to those observed in experiments and in DNS of incompressible flows \cite{hof_science_2004,kerswell_jfm_2007}.
 }
\end{figure*}

DPD is a well-tested and documented \cite{hoogerbrugge_epl_1992,groot_jcp_1997} off-lattice method to simulate the Navier-Stokes equations. Its popularity is partly due to the ease of extending it to multiphase and viscoelastic flows. A limitation of the method which is not often stressed but which will come to the foreground here is that particle interactions have intrinsic timescales such that a DPD fluid is highly compressible.

For plane Couette and pipe flows we find that at large enough $\Re$, there is a hysteretic transition to a weakly-turbulent state dominated by coherent structures much like those present in DNS and experiments --- see Figs.\ref{fig:couette_2d} and \ref{fig:pipe_2d}. As the compressibility increases, the transition to turbulence shifts to higher Reynolds numbers and becomes less abrupt and possibly even continuous, but the overall features in our fluid with high compressibility and strong thermal fluctuations are similar to those of incompressible fluids without thermal fluctuations.

{\em DPD simulation method --- } In DPD one integrates Newton's equations for a system of unit-mass particles that represent \emph{parcels} of fluid. Forces between particles  are chosen so as to optimize the convergence to hydrodynamic behavior on length scales of a few particles --- a rationale that is similar to that of the Lattice-Boltzmann method \cite{succi_2001}, where microscopic fidelity is also sacrificed to obtain a computationally efficient representation of hydrodynamics beyond the lattice scale. The DPD interparticle forces are pair-wise and consist of three contributions: a soft-repulsion conservative component $f_{c}  = a \: (1-r)$, a dissipative component $f_{d} = -\g \: (1-r)^{1/2} \: ( \vec{\delta v} \cdot \hat{r} )$ that tends to reduce the difference in particle velocities $\vec{\delta v}$, and a stochastic component $f_{r} = \s \: (1-r)^{1/4} \chi (t)$. The constants $a$, $\g$ and $\s$ define the amplitude of each of the components; $r$ is the distance between the particles and $\hat{r}$ is the unit vector in the direction of $\vec{r}$. The range of interaction is customarily set to 1. A Gaussian-distributed random variable $\chi$ with unit variance defines the evolution of random interactions. The amplitude $\s$ and the form of the dissipative and random forces are chosen so that a DPD fluid at rest satisfies the fluctuation-dissipation theorem with temperature $T$:  $\s^2 = (2 \g kT/ \D t)$. The absorption of a factor $\sqrt{1/\D t}$  into $\s$ is necessary to  converge properly to a continuum limit for small timesteps  \cite{groot_jcp_1997}.
An important though seldom stressed point of  DPD is that in order to converge quickly to hydrodynamic length-scales upon coarse graining, the three force components have to be of roughly the same size. This effectively limits the parameters $(a,\g,kT)$ to the range 0.1-10 in DPD units, and implies that   density fluctuations, viscous interaction and thermal diffusion all take place on similar timescales. On the particle scale, DPD thus  models a compressible and hot sluggish fluid. In our simulations, we focus on the effect of the compressibility on the transition to turbulence. One should keep in mind, however, that the thermal fluctuations can also be
important --- even though the thermal velocities $v_{\mathrm {th}}$ are much smaller than the flow velocity scale $U$ (in our case typically $v_{\mathrm{ th}}/U \sim 10^{-2}$), they determine a  natural Reynolds number above which the flow
will become unstable even without external perturbations, because the thermal fluctuations are sufficient to destabilize the flow due to the subcritical nature of the instability.

{\em Flow domains --- }
Simulations were carried out in two classical geometries: flow between two plates $y=\pm h$ sliding with  opposite velocities $\pm U$ along each other (plane Couette flow), and pressure-driven flow in a pipe created by applying  a constant force in the flow direction to \emph{every} particle. The simulation box for the plane Couette geometry has dimensions $60 \times 20 \times 40$ in the streamwise (velocity) $x$, gradient direction $y$, and spanwise direction $z$, and is periodic in the $x$ and $z$ directions. The DPD parameters are $(a,\g,kT)=(8,1,0.35)$ unless stated otherwise. The density is $\rho=4$, corresponding to $N=192000$ particles in the system. The parameter values for pipe flow are similar and given in the caption of Fig.~\ref{fig:pipe_2d}.

To impose the no-slip boundary conditions at the walls, we employ the method introduced by Revenga {\em et al.} \cite{revenga_ijmpc_1998,revenga_cpc_1999} for 2D simulations. Here, the walls are modelled by an immobile continuum DPD medium of uniform density which interacts with the bulk particles (see \cite{vandemeent_2006} for details). We have checked that in all our runs the velocity difference between a wall and the first layer of particles close to the wall is indeed negligibly small.

{\em Reynolds and Mach numbers ---} 
In our simulations, we characterize the flow by two parameters: the Reynolds number, defined as $\Re=hU/\nu$ for Couette flow (for pipe flow $\Re=dU/\nu$, where $U$ is the \emph{averaged} streamwise velocity and $d$ the pipe diameter), and the Mach number $\Ma = U/c $, where $c$ is the sound velocity. We empirically obtain $c$ by measuring the speed of propagation of a density pulse directly, yielding $c=4.4$ for the  parameter set mentioned above. The  sound velocity can also be estimated by $c \approx \sqrt{\partial p / \partial \rho |_T}$ based on  the virial expansion for the pressure $p(\rho,T)$, which typically yields estimates within 20\% of our direct measurements. The kinematic viscosity was measured from the laminar shear start-up and was found to be $\nu=0.23$. Since  we vary $U$ keeping other parameters fixed, we report our data  in terms of the ratio $\Re/ \Ma= hc/\nu$, which is of order 90-220 ($\Ma\simeq 2-5$). Thus, we are sampling a different parameter range than accessible in experiments \cite{hof_science_2004,hof_prl_2005} and theory \cite{nagata_jfm_1990,waleffe_prl_1998,waleffe_pf_2003,faisst_prl_2003,wedin_jfm_2004,wang_prl_2007,skufca_prl_2006,kerswell_jfm_2007,viswanath_archive_2007,gibson_archive_2007}.

\begin{figure}[!t]
  \includegraphics[width=2.6in]{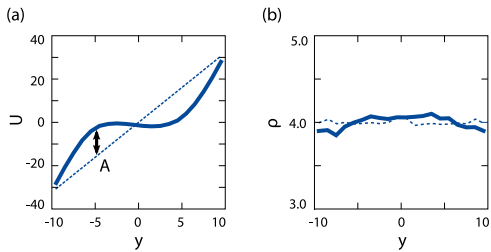}
   \vspace*{-3mm}
  
  \caption{\label{fig:couette_profiles} Mean streamwise velocity $U(y)$ and density $\r(y)$ for the flow snapshot shown in Fig.~\ref{fig:couette_2d} (solid), with the initial $t=0$ state (dotted) shown for comparison. (\emph{a}) The mean velocity $U(y)$ develops a sinusoidal deviation from the linear profile as coherent structures develop.  (\emph{b}) As a result of the compressibility of the fluid, the density profile $\rho(y)$ shows a slight bulge at the center of the  cell.}
\end{figure}

{\em Results for plane Couette flow ---}
To quantify the influence of the compressibility on the transition to turbulence, we focus on the plane Couette geometry. The qualitative features of the velocity field are in agreement with the SSP predictions: Fig.~1a shows the streamwise vortices and streaks at $\Re=1400$. Moreover, as Fig.~1b shows, the streaks have a slight sinusoidal modulation in the streamwise-spanwise plane, in agreement with the SSP scenario that a Kelvin-Helmholtz-type instability transfers energy back into the vortices. Finally, the mean velocity profile of Fig.~\ref{fig:couette_profiles}a  strongly deviates from the laminar profile in the way typical for turbulent flow. The mean density of the system, plotted in Fig.~\ref{fig:couette_profiles}b,  deviates slightly from its equilibrium value, which emphasizes the compressible nature of our fluid.

The streaky profile in plane Couette flow tends to have well-defined modes exhibiting an integer number of streak and vortex pairs that are fairly persistent in time. The bifurcation point and the dominant threshold mode depend on the Mach number. For low $\Re/\Ma$, the first mode appearing is typically a one-pair streak-vortex configuration as shown in Fig.~\ref{fig:couette_2d}, while for higher $\Re/\Ma$ our simulations exhibit a bifurcation towards a two-pair configuration. Higher order configurations appear as the Reynolds number is increased further and switching between configurations becomes more frequent.

\begin{figure}[!t]
 \includegraphics[width=3in]{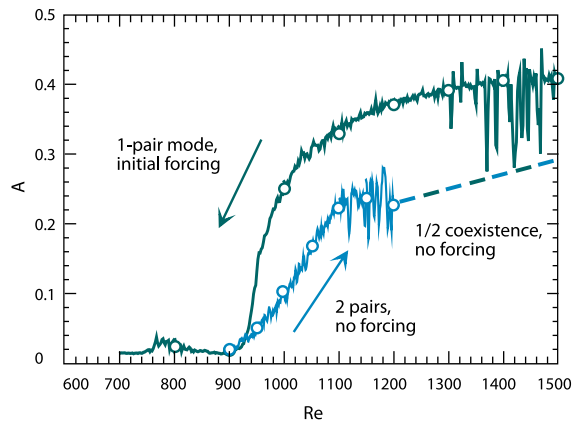}
   \vspace*{-3mm}
  
 \caption{\label{fig:couette_bifurcation} Bistability in the bifurcation from laminar flow at $\Re/\Ma=190$. Shown is the maximum deviation $A$ of the mean profile $U(y)$ from linearity, $A =\mathrm{max_y\:} |(U(y)/U \!-\! y/h)|$. Approaching from the laminar state, a smooth (\emph{forward}) transition towards a 2-pair mode is observed, which becomes unstable at $Re\gtrsim 1200$. 
 A stable 1-pair mode is found by initial forcing of a pair of rolls. Decreasing the driving rate results in a jump (\emph{subcritical}) transition back to the laminar state. }
\end{figure}

Reproducibility of the dynamics observed close to the threshold allows examination of the transition in terms of the bifurcation diagram of Fig.~\ref{fig:couette_bifurcation}. We plot the deviation $A$ of the normalised profile $U(y)/U$ from the laminar flow for a series of states initialized at regular intervals in $\Re$, marked with circles. The \emph{lower curve} denotes the 2-pair modes which bifurcate spontaneously from the laminar state. The \emph{upper curve} denotes 1-pair modes that are created by initially driving all the particles with an additional external force term with the desired symmetry. The vortices that are thus created either persist or relaminarize after the forcing is turned off. The open circles on this branch denote states obtained this way. On both branches, the states can be traced smoothly by adiabatically increasing/decreasing the driving rate, thereby producing the continuous curves. The fact that the segments connect perfectly shows that the amplitude of turbulent perturbations is a well-defined function of $\Re$ and that the adjustment of $U$ can be considered adiabatic. 

Contrary to the results for incompressible flows, the bifurcation observed in the lower curve of Fig.~\ref{fig:couette_bifurcation} appears continuous in $\Re$ rather than  jump-like. At this point we are not sure which property  underlies this qualitative difference in the onset dynamics --- compressibility,  finite temperature $T$ or finite size. One possibility is that the 2-pair mode has an instability threshold smaller than the typical fluctuating velocities while the resulting nonlinear branch lies so low that the transition appears continuous in $\Re$. At the same time, the forced 1-pair state has presumably a higher instability threshold (since its upper branch is higher than the upper branch of the 2-pair mode) and when tracked down along the upper branch, still exhibits a jump back to the laminar flow.

\begin{figure}[!t]
 \includegraphics[width=3in]{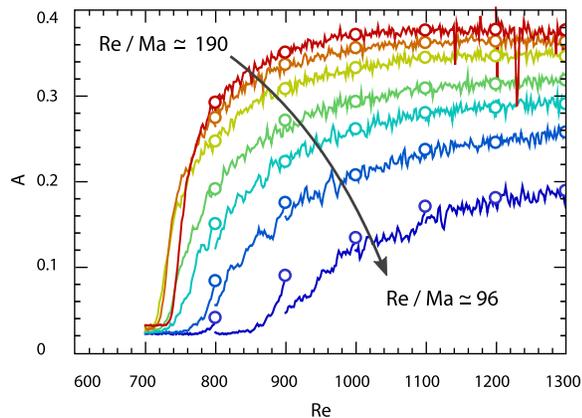}
   \vspace*{-3.5mm}
  
 \caption{\label{fig:couette_nose_shift} Relaminarization in a series of runs with increasing compressibility. The box size is $48 \times 16 \times 32$. One-pair modes are initialized at regular intervals in $\Re$, with the sound velocities ranging from $c \simeq 5.5$ (red) to $c \simeq 2.8$ (blue) and tracked down by slowly decreasing the driving rate. The resulting curves show that  the compressibility decreases the amplitude of the turbulent state and the abruptness of the transition.}
\end{figure}

We now proceed to examine how these features change when we change the compressibility and hence the Mach number $\Ma$ by tuning the repulsive force strength $a$. Fig.~\ref{fig:couette_nose_shift} shows 7 upper branches for a range of $\Ma$. The initial states are created by seeding 1-pair modes at regularly spaced intervals in $\Re$, followed by a slow adjustment of the conservative force strength $a$ towards the values $a=3,4,5,6,8,10,13$. The corresponding sound velocities are $c \simeq 2.8,3.2,3.5,3.8,4.4,4.9,5.5$, and so for fixed driving velocity the Mach numbers of these runs varies by a factor of 2. Subsequent interpolation of the states by adiabatic adjustment of the driving rate then results in the curves shown in the figure. Clearly there is an unmistakable suppression of the turbulence as the compressibility increases: the turbulent amplitude decreases and the onset shifts to larger $\Re$. In addition the bistable jump behavior changes gradually into an apparently smooth transition as the compressibility increases. 

{\em Results for pipe flow --- } Fig.~2 shows snapshots of our results for pipe flow for two Reynolds numbers $\Re=1700$ and $\Re=2700$. These results are consistent with the SSP scenario and are similar to those found in experiments and simulations of incompressible flows \cite{hof_science_2004,hof_prl_2005,kerswell_jfm_2007,kerswell_nonlinearity_2005,eckhardt_arfm_2007}. They also resemble exact solutions of the incompressible Navier-Stokes equations \cite{faisst_prl_2003,wedin_jfm_2004}. Streaks, visible as colored contours where the downstream velocity is higher (red) or lower (blue) than average, are stabilized by the streamwise vortices (vectors). As in incompressible flows \cite{kerswell_jfm_2007}, we observe dynamical transitions between these states. We leave a detailed study of this process for the future.

{\em Conclusion ---} DPD simulations reproduce the qualitative features of the exact coherent structures in remarkable detail. Considering the large degree of compressibility in DPD fluids, the fact that the phenomenology differs so little constitutes new support for the SSP  \cite{waleffe_pf_1997} as  {\em the} scenario for organization of turbulence at moderate Reynolds numbers. The dependence of turbulence amplitudes on $\Ma$ in the DPD fluid provides clear evidence for a suppressing effect of the compressibility on the transition to turbulence, with apparently a crossover to a continuous transition in a fluctuating fluid.

Further potential of the method lies in its flexibility in incorporating interactions between fluid elements. DPD is very easy to program, and our simulations with over $10^5 $ particles are quite feasible on a single node computer. Turbulence in multiphase and viscoelastic flows \cite{somfai_pa_2006}, and in other complex fluids, clearly seems within reach.

We thank Bruno Eckhardt and Pep Espa\~{n}ol for valuable discussions, the EU network PHYNECS and Dutch science foundations NWO and FOM for support, and the national computer center SARA for computer time.

\bibliography{journals_abbr_ams,dpd,turbulence}

\begin{thebibliography}{25}
\expandafter\ifx\csname natexlab\endcsname\relax\def\natexlab#1{#1}\fi
\expandafter\ifx\csname bibnamefont\endcsname\relax
  \def\bibnamefont#1{#1}\fi
\expandafter\ifx\csname bibfnamefont\endcsname\relax
  \def\bibfnamefont#1{#1}\fi
\expandafter\ifx\csname citenamefont\endcsname\relax
  \def\citenamefont#1{#1}\fi
\expandafter\ifx\csname url\endcsname\relax
  \def\url#1{\texttt{#1}}\fi
\expandafter\ifx\csname urlprefix\endcsname\relax\def\urlprefix{URL }\fi
\providecommand{\bibinfo}[2]{#2}
\providecommand{\eprint}[2][]{\url{#2}}

\bibitem[{\citenamefont{Nagata}(1990)}]{nagata_jfm_1990}
\bibinfo{author}{\bibfnamefont{M.}~\bibnamefont{Nagata}}, \bibinfo{journal}{J.
  Fluid Mech.} \textbf{\bibinfo{volume}{217}}, \bibinfo{pages}{519}
  (\bibinfo{year}{1990}).

\bibitem[{\citenamefont{Waleffe}(1998)}]{waleffe_prl_1998}
\bibinfo{author}{\bibfnamefont{F.}~\bibnamefont{Waleffe}},
  \bibinfo{journal}{Phys. Rev. Lett.} \textbf{\bibinfo{volume}{81}},
  \bibinfo{pages}{4140} (\bibinfo{year}{1998}).

\bibitem[{\citenamefont{Waleffe}(2003)}]{waleffe_pf_2003}
\bibinfo{author}{\bibfnamefont{F.}~\bibnamefont{Waleffe}},
  \bibinfo{journal}{Phys. Fluids} \textbf{\bibinfo{volume}{15}},
  \bibinfo{pages}{1517} (\bibinfo{year}{2003}).

\bibitem[{\citenamefont{Faisst and Eckhardt}(2003)}]{faisst_prl_2003}
\bibinfo{author}{\bibfnamefont{H.}~\bibnamefont{Faisst}} \bibnamefont{and}
  \bibinfo{author}{\bibfnamefont{B.}~\bibnamefont{Eckhardt}},
  \bibinfo{journal}{Phys. Rev. Lett.} \textbf{\bibinfo{volume}{91}},
  \bibinfo{pages}{224502} (\bibinfo{year}{2003}).

\bibitem[{\citenamefont{Wedin and Kerswell}(2004)}]{wedin_jfm_2004}
\bibinfo{author}{\bibfnamefont{H.}~\bibnamefont{Wedin}} \bibnamefont{and}
  \bibinfo{author}{\bibfnamefont{R.~R.} \bibnamefont{Kerswell}},
  \bibinfo{journal}{J. Fluid Mech.} \textbf{\bibinfo{volume}{508}},
  \bibinfo{pages}{333} (\bibinfo{year}{2004}).

\bibitem[{\citenamefont{Wang et~al.}(2007)\citenamefont{Wang, Gibson, and
  Waleffe}}]{wang_prl_2007}
\bibinfo{author}{\bibfnamefont{J.}~\bibnamefont{Wang}},
  \bibinfo{author}{\bibfnamefont{J.}~\bibnamefont{Gibson}}, \bibnamefont{and}
  \bibinfo{author}{\bibfnamefont{F.}~\bibnamefont{Waleffe}},
  \bibinfo{journal}{Phys. Rev. Lett.} \textbf{\bibinfo{volume}{98}},
  \bibinfo{pages}{204501} (\bibinfo{year}{2007}).

\bibitem[{\citenamefont{Robinson}(1991)}]{robinson_arfm_1991}
\bibinfo{author}{\bibfnamefont{S.~K.} \bibnamefont{Robinson}},
  \bibinfo{journal}{Annu. Rev. Fluid Mech.} \textbf{\bibinfo{volume}{23}},
  \bibinfo{pages}{601} (\bibinfo{year}{1991}).

\bibitem[{\citenamefont{Waleffe}(1997)}]{waleffe_pf_1997}
\bibinfo{author}{\bibfnamefont{F.}~\bibnamefont{Waleffe}},
  \bibinfo{journal}{Phys. Fluids} \textbf{\bibinfo{volume}{9}},
  \bibinfo{pages}{883} (\bibinfo{year}{1997}).

\bibitem[{\citenamefont{Hof et~al.}(2004)\citenamefont{Hof, van Doorne,
  Westerweel, Nieuwstadt, Faisst, and Eckhardt}}]{hof_science_2004}
\bibinfo{author}{\bibfnamefont{B.}~\bibnamefont{Hof}},
  \bibinfo{author}{\bibfnamefont{C.~W.~H.} \bibnamefont{van Doorne}},
  \bibinfo{author}{\bibfnamefont{J.}~\bibnamefont{Westerweel}},
  \bibinfo{author}{\bibfnamefont{F.~T.~M.} \bibnamefont{Nieuwstadt}},
  \bibinfo{author}{\bibfnamefont{H.}~\bibnamefont{Faisst}}, \bibnamefont{and}
  \bibinfo{author}{\bibfnamefont{B.}~\bibnamefont{Eckhardt}},
  \bibinfo{journal}{Science} \textbf{\bibinfo{volume}{305}},
  \bibinfo{pages}{1594} (\bibinfo{year}{2004}).

\bibitem[{\citenamefont{Hof et~al.}(2005)\citenamefont{Hof, van Doorne,
  Westerweel, and Nieuwstadt}}]{hof_prl_2005}
\bibinfo{author}{\bibfnamefont{B.}~\bibnamefont{Hof}},
  \bibinfo{author}{\bibfnamefont{C.~W.~H.} \bibnamefont{van Doorne}},
  \bibinfo{author}{\bibfnamefont{J.}~\bibnamefont{Westerweel}},
  \bibnamefont{and} \bibinfo{author}{\bibfnamefont{F.~T.~M.}
  \bibnamefont{Nieuwstadt}}, \bibinfo{journal}{Phys. Rev. Lett.}
  \textbf{\bibinfo{volume}{95}}, \bibinfo{pages}{214502}
  (\bibinfo{year}{2005}).

\bibitem[{\citenamefont{Skufca et~al.}(2006)\citenamefont{Skufca, Yorke, and
  Eckhardt}}]{skufca_prl_2006}
\bibinfo{author}{\bibfnamefont{J.~D.} \bibnamefont{Skufca}},
  \bibinfo{author}{\bibfnamefont{J.~A.} \bibnamefont{Yorke}}, \bibnamefont{and}
  \bibinfo{author}{\bibfnamefont{B.}~\bibnamefont{Eckhardt}},
  \bibinfo{journal}{Phys. Rev. Lett.} \textbf{\bibinfo{volume}{96}},
  \bibinfo{pages}{174101} (\bibinfo{year}{2006}).

\bibitem[{\citenamefont{Kerswell and Tutty}(2007)}]{kerswell_jfm_2007}
\bibinfo{author}{\bibfnamefont{R.~R.} \bibnamefont{Kerswell}} \bibnamefont{and}
  \bibinfo{author}{\bibfnamefont{O.~R.} \bibnamefont{Tutty}},
  \bibinfo{journal}{J. Fluid Mech.} \textbf{\bibinfo{volume}{584}},
  \bibinfo{pages}{69} (\bibinfo{year}{2007}).

\bibitem[{\citenamefont{Viswanath}(2007)}]{viswanath_archive_2007}
\bibinfo{author}{\bibfnamefont{D.}~\bibnamefont{Viswanath}}
  (\bibinfo{year}{2007}), \bibinfo{note}{arXiv:physics/0701337}.

\bibitem[{\citenamefont{Gibson et~al.}(2007)\citenamefont{Gibson, Halcrow, and
  Cvitanovi\'{c}}}]{gibson_archive_2007}
\bibinfo{author}{\bibfnamefont{J.~F.} \bibnamefont{Gibson}},
  \bibinfo{author}{\bibfnamefont{J.}~\bibnamefont{Halcrow}}, \bibnamefont{and}
  \bibinfo{author}{\bibfnamefont{P.}~\bibnamefont{Cvitanovi\'{c}}}
  (\bibinfo{year}{2007}), \bibinfo{note}{arXiv:physics/07053957}.

\bibitem[{\citenamefont{Kerswell}(2005)}]{kerswell_nonlinearity_2005}
\bibinfo{author}{\bibfnamefont{R.~R.} \bibnamefont{Kerswell}},
  \bibinfo{journal}{Nonlinearity} \textbf{\bibinfo{volume}{18}},
  \bibinfo{pages}{17} (\bibinfo{year}{2005}).

\bibitem[{\citenamefont{Eckhardt et~al.}(2007)\citenamefont{Eckhardt,
  Schneider, Hof, and Westerweel}}]{eckhardt_arfm_2007}
\bibinfo{author}{\bibfnamefont{B.}~\bibnamefont{Eckhardt}},
  \bibinfo{author}{\bibfnamefont{T.~M.} \bibnamefont{Schneider}},
  \bibinfo{author}{\bibfnamefont{B.}~\bibnamefont{Hof}}, \bibnamefont{and}
  \bibinfo{author}{\bibfnamefont{J.}~\bibnamefont{Westerweel}},
  \bibinfo{journal}{Annu. Rev. Fluid Mech.} \textbf{\bibinfo{volume}{39}},
  \bibinfo{pages}{447} (\bibinfo{year}{2007}).

\bibitem[{\citenamefont{Moehlis et~al.}(2004)\citenamefont{Moehlis, Faisst, and
  Eckhardt}}]{moehlis_njp_2004}
\bibinfo{author}{\bibfnamefont{J.}~\bibnamefont{Moehlis}},
  \bibinfo{author}{\bibfnamefont{H.}~\bibnamefont{Faisst}}, \bibnamefont{and}
  \bibinfo{author}{\bibfnamefont{B.}~\bibnamefont{Eckhardt}},
  \bibinfo{journal}{New J. Phys.} \textbf{\bibinfo{volume}{6}},
  \bibinfo{pages}{56} (\bibinfo{year}{2004}).

\bibitem[{\citenamefont{Hoogerbrugge and
  Koelman}(1992)}]{hoogerbrugge_epl_1992}
\bibinfo{author}{\bibfnamefont{P.~J.} \bibnamefont{Hoogerbrugge}}
  \bibnamefont{and} \bibinfo{author}{\bibfnamefont{J.}~\bibnamefont{Koelman}},
  \bibinfo{journal}{Europhys. Lett.} \textbf{\bibinfo{volume}{19}},
  \bibinfo{pages}{155} (\bibinfo{year}{1992}).

\bibitem[{\citenamefont{Groot and Warren}(1997)}]{groot_jcp_1997}
\bibinfo{author}{\bibfnamefont{R.~D.} \bibnamefont{Groot}} \bibnamefont{and}
  \bibinfo{author}{\bibfnamefont{P.~B.} \bibnamefont{Warren}},
  \bibinfo{journal}{J. Chem. Phys.} \textbf{\bibinfo{volume}{107}},
  \bibinfo{pages}{4423} (\bibinfo{year}{1997}).

\bibitem[{\citenamefont{Succi}(2001)}]{succi_2001}
\bibinfo{author}{\bibfnamefont{S.}~\bibnamefont{Succi}},
  \emph{\bibinfo{title}{The Lattice Boltmann Equation}}
  (\bibinfo{publisher}{Clarendon Press}, \bibinfo{address}{Oxford},
  \bibinfo{year}{2001}).

\bibitem[{\citenamefont{Chen et~al.}(2003)\citenamefont{Chen, Kandasamy,
  Orszag, Shock, Succi, and Yakhot}}]{chen_sci_2003}
\bibinfo{author}{\bibfnamefont{H.~D.} \bibnamefont{Chen}},
  \bibinfo{author}{\bibfnamefont{S.}~\bibnamefont{Kandasamy}},
  \bibinfo{author}{\bibfnamefont{S.}~\bibnamefont{Orszag}},
  \bibinfo{author}{\bibfnamefont{R.}~\bibnamefont{Shock}},
  \bibinfo{author}{\bibfnamefont{S.}~\bibnamefont{Succi}}, \bibnamefont{and}
  \bibinfo{author}{\bibfnamefont{V.}~\bibnamefont{Yakhot}},
  \bibinfo{journal}{Science} \textbf{\bibinfo{volume}{301}},
  \bibinfo{pages}{633} (\bibinfo{year}{2003}).

\bibitem[{\citenamefont{Revenga et~al.}(1998)\citenamefont{Revenga,
  Z{\'u}{\~n}iga, Espa{\~n}ol, and Pagonabarraga}}]{revenga_ijmpc_1998}
\bibinfo{author}{\bibfnamefont{M.}~\bibnamefont{Revenga}},
  \bibinfo{author}{\bibfnamefont{I.}~\bibnamefont{Z{\'u}{\~n}iga}},
  \bibinfo{author}{\bibfnamefont{P.}~\bibnamefont{Espa{\~n}ol}},
  \bibnamefont{and}
  \bibinfo{author}{\bibfnamefont{I.}~\bibnamefont{Pagonabarraga}},
  \bibinfo{journal}{Int. J. Mod. Phys. C} \textbf{\bibinfo{volume}{9}},
  \bibinfo{pages}{1319} (\bibinfo{year}{1998}).

\bibitem[{\citenamefont{Revenga et~al.}(1999)\citenamefont{Revenga, Zuniga, and
  Espanol}}]{revenga_cpc_1999}
\bibinfo{author}{\bibfnamefont{M.}~\bibnamefont{Revenga}},
  \bibinfo{author}{\bibfnamefont{I.}~\bibnamefont{Zuniga}}, \bibnamefont{and}
  \bibinfo{author}{\bibfnamefont{P.}~\bibnamefont{Espanol}},
  \bibinfo{journal}{Comp. Phys. Commun.} \textbf{\bibinfo{volume}{121}},
  \bibinfo{pages}{309} (\bibinfo{year}{1999}).

\bibitem[{\citenamefont{{van de Meent}}(2006)}]{vandemeent_2006}
\bibinfo{author}{\bibfnamefont{J.~W.} \bibnamefont{{van de Meent}}}, Master's
  thesis, \bibinfo{school}{Universiteit Leiden} (\bibinfo{year}{2006}),
  \urlprefix\url{http://www. ilorentz. org/~saarloos/thesis_jw_vdmeent_2006.
  pdf}.

\bibitem[{\citenamefont{Somfai et~al.}(2006)\citenamefont{Somfai, Morozov, and
  van Saarloos}}]{somfai_pa_2006}
\bibinfo{author}{\bibfnamefont{E.}~\bibnamefont{Somfai}},
  \bibinfo{author}{\bibfnamefont{A.~N.} \bibnamefont{Morozov}},
  \bibnamefont{and} \bibinfo{author}{\bibfnamefont{W.}~\bibnamefont{van
  Saarloos}}, \bibinfo{journal}{Physica A} \textbf{\bibinfo{volume}{362}},
  \bibinfo{pages}{93} (\bibinfo{year}{2006}).

\end{thebibliography}

\end{document}